\documentclass[%
reprint,
amsmath,amssymb,
aip,
]{revtex4-2}

\usepackage{graphicx}
\usepackage{dcolumn}
\usepackage{bm}
\usepackage{xcolor}

\newcommand{\bvec}{\begin{pmatrix}}
\newcommand{\evec}{\end{pmatrix}}

\newcommand{\D}{\mathrm{d}}

\graphicspath{{figs}}

\begin{document}
	
	
	\title{Bremsstrahlung constraints on proton-Boron 11 inertial fusion}

\author{Ian E. Ochs}
\email{iochs@princeton.edu}
\author{Elijah J. Kolmes}
\author{Alexander S. Glasser}
\author{Nathaniel J. Fisch}%
\affiliation{Department of Astrophysical Sciences, Princeton University, Princeton, NJ 08540}%

\date{\today}

\begin{abstract}
	Proton-Boron 11 (pB11) fusion is relatively safe and clean, but difficult to use for net power production, since bremsstrahlung radiation tends to radiate away power more quickly than it can be generated by fusion power, particularly once poisoning by alpha particles is taken into account.
	While in magnetic confinement fusion (MCF), this problem can be addressed by deconfining the alphas, in inertial confinement fusion (ICF) the alphas that heat the plasma linger for the duration of the reaction.
	Thus, it becomes essential to trap the bremsstrahlung radiation in the hotspot.
	Through burn simulations incorporating bremsstrahlung emission and reabsorption, we infer the necessary conditions to capture enough radiation to produce scientific breakeven in a pB11 ICF plasma.
    We find that breakeven requires a stagnation areal density roughly two orders of magnitude higher than the current state-of-the-art, at pressures three orders of magnitude higher.
\end{abstract}

\maketitle


\section{Introduction} 
Proton-boron 11 (pB11) fusion is appealing due to its lack of neutron production and its abundant and safe reactants and byproducts.
While net energy production from pB11 was long dismissed as impossible, \cite{Rider1995FundamentalLimitations,Rider1995GeneralCritique} partly due to erroneously low cross-section data, \cite{Nevins2000ThermonuclearFusion}
newer cross-section data \cite{Sikora2016NewEvaluation} has opened up a broader range of feasibility for both pB11 ignition \cite{Putvinski2019FusionReactivity,Ochs2022ImprovingFeasibility,Kolmes2022WavesupportedHybrid} and net energy production \cite{Ochs2024LoweringReactor}.
,Partly as a result, there has been an explosion of both public and private sector interest in pB11 fusion. \cite{Rostoker1997CollidingBeam,Lampe1998NRLMR,Volosov2006AneutronicFusion,Volosov2011ProblemsACT,Labaune2013FusionReactions,Eliezer2016AvalancheProtonboron,Magee2019DirectObservation,Eliezer2020NovelFusion,Eliezer2020MitigationStopping,Ruhl2022NonthermalLaserdriven,Istokskaia2023MultiMeVAlpha,Wei2023ProtonBoronFusion,Magee2023FirstMeasurements,Liu2024ENNsRoadmap,Xie2025PreliminaryConsiderations,Liu2025FeasibilityProton,Hwang2025UpperLimit}

The inertial confinement fusion (ICF) approach to pB11 has both advantages and disadvantages over the magnetic confinement fusion (MCF) approach. 
One disadvantage is ash handling. 
The pB11 reaction produces 3 $\alpha$ particles (He nuclei), which initially contain the 8.7 MeV of energy released by the fusion reaction.
Previously, it was shown that in a steady-state plasma, this helium ash naturally poisons the reaction, causing the bremsstrahlung radiation to dramatically exceed the fusion power and thus prevent breakeven power production even with highly efficient heating and power recovery systems.\cite{Ochs2025PreventingAsh}
Thus, it becomes necessary to remove the helium on a timescale shorter than the energy confinement time, e.g. through alpha channeling \cite{Fisch1992InteractionEnergetic,Fisch1995AlphaPowera} or manipulation of the ion transport,\cite{Kolmes2018StrategiesAdvantageous} so as to return to the more optimistic scenarios without alpha poisoning.\cite{Putvinski2019FusionReactivity, Ochs2022ImprovingFeasibility,Kolmes2022WavesupportedHybrid,Ochs2024LoweringReactor}

In MCF plasmas, such manipulations of the helium ash are in theory possible, as a variety of laboratory-scale plasma control tools can be brought to bear.
These techniques are not so readily applied to ICF plasmas, where reactions occur on length scales of microns and timescales of nanoseconds.

ICF also comes with certain dynamical challenges that do not appear in MCF. 
In particular, efficient power generation requires ICF plasmas to burn up a reasonable fraction of their fuel with each shot. 
This challenge is particularly acute for pB11 ICF, as the fuel reactivity is generally lower than that of the deuterium-tritium reaction. 

However, ICF plasmas have an advantage over MCF plasmas when it comes to radiative losses: namely, they are incredibly dense. 
This density allows the possibility of inverse bremsstrahlung absorption, allowing the plasma to recapture bremsstrahlung energy before it leaves, and potentially allowing the fusion power to exceed the escaping bremsstrahlung power. 
This is a potentially very significant upside: the bremsstrahlung loss channel may be the single most significant reason for the perceived difficulty of pB11 power generation, and it is difficult to mitigate in the MCF regime. 
The aim of this work is to show, under a very simplified set of assumptions, that bremsstralung must be trapped to achieve reasonable gain in ICF pB11 scenarios, and that this cannot be achieved for realistic near-term parameters barring some unusual intervention.  
Among other simplifications, our model assumes classical plasma near fusion temperatures, neglects ion and electron heat transport, and neglects the evolution of plasma density during the disassembly of the fuel. These very significant  simplifications isolate the key limitations in pB11 inertial fusion approaches while considering a wide array of possible regimes, thereby better to point to  the regimes that might be most susceptible to the overcoming of these limitations.

To perform this analysis, we begin in Sec.~\ref{sec:heuristics} by very approximately estimating the performance of pB11 ICF fusion plasma that is transparent to bremsstrahlung, showing how the requirement of sustained plasma burn makes a breakeven fusion reaction basically impossible.
We then deepen our analysis in Sec.~\ref{sec:BurnModel} by deriving a rate equation model for the plasma burn at stagnation, assuming all bremsstrahlung radiation is lost.
In Sec.~\ref{sec:QTransparent}, we present the results of these simulations. 
By evaluating the initial invested energy and the burn fraction, we determine the maximum achievable $Q_\text{sci}$ (ratio of fusion energy to assembly energy) as a function of $f_b$, $T_i$, and $\bar{T}_e$, showing that $Q_\text{sci} > 3$ is unachievable in an ICF pB11 plasma transparent to bremsstrahlung. 
Thus, in Sec.~\ref{sec:BremsReabsorption}, we introduce a model for bremsstrahlung reabsorption, and incorporate this model into the rate equation model.
In Sec.~\ref{sec:QWithReabsorption}, we present the results of simulations with reabsorption showing how the reabsorption allows for the presence of much higher-performance plasmas consistent with a power plant.
In Sec.~\ref{sec:Yield}, we combine the necessities for bremsstrahlung reabsorption and tolerable (non-bomb) reaction yields to determine the parameter space for an ICF power plant, showing that it requires particle densities on the order of $10^{28}$ cm$^{-3}$ and areal densities on the order of 20 g/cm$^2$; approximately two orders of magnitude above current NIF parameters, and thus essentially unrealistic in the near future. 
Finally, in Sec.~\ref{sec:Discussion}, we discuss possible ways to navigate around these constraints, as well as possible shortcomings with the model (including the lack of electron degeneracy) that should be addressed in future work.

\section{The need for high burn fraction and bremsstrahlung capture}  \label{sec:heuristics}

To achieve a high-performing ICF plasma, the first challenge is to produce sustained burn,\cite{Hurricane2019ApproachingBurning}
which requires that the fusion power be sufficient to balance against cooling from electron heat conduction and bremsstrahlung losses. 
In this paper, we will ignore the (often significant) electron heat transport terms, and focus solely on irreducible bremsstrahlung radiation, focusing on the bremsstrahlung.

\begin{figure}
	\centering
	\includegraphics[width=\linewidth]{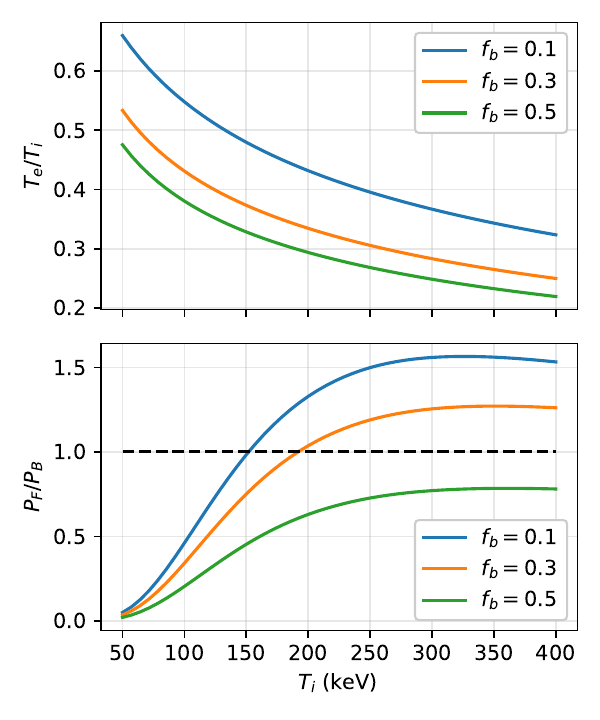}
	\caption{Steady-state electron temperature from Eq.~(\ref{eq:ElectronQuasiSteadyState}) as a function of ion temperature (top), and the resulting ratio between fusion power and bremsstrahlung power given those electron and ion temperatures (bottom), for several values of the boron fraction $f_b \equiv n_b/n_i$.
		With too high a boron fraction, the fusion power cannot exceed the bremsstrahlung power; thus, a boron fraction less than 50\% is necessary to achieve sustained burn, with lower $f_b$ allowing burn at lower ion temperatures.}
	\label{fig:TePbPf_Vs_Ti}
\end{figure}

The requirement for sustained burn sets a minimal initial ion temperature for the stagnating fusion plasma.
To see how this minimal temperature emerges, consider the electron energy balance equation in steady state.
Electrons lose energy through bremsstrahlung $P_B$, and gain energy through collisions $K_{ei}$ with ions:
\begin{align}
	\frac{dU_e}{dt} = - P_B +  K_{ei}  (T_i - T_e) = 0. \label{eq:ElectronQuasiSteadyState}
\end{align}
The formulae for calculating $P_B$ and $K_{ei}$ are descibed later, in Sec.~\ref{sec:BurnModel}.
Using Eq.~(\ref{eq:ElectronQuasiSteadyState}), we can estimate the electron temperature given the ion temperature.
From this temperature we can then compare the fusion power $P_F$ and bremsstrahlung power $P_B$, for a given ion temperature $T_i$ and boron fraction $f_b$.
The results are shown in Fig.~\ref{fig:TePbPf_Vs_Ti}.

We can see two distinct trends emerge.
First, because of boron's higher charge state, a higher boron fraction results in a higher ratio of bremsstrahlung power to fusion power, making sustained burn more difficult to achieve.
For $f_b = 0.1$, the fusion power exceeds the bremsstrahlung power at ion temperatures above 150 keV; for $f_b = 0.3$, around 200 keV; and for $f_b = 0.5$, never.
Thus, the need for burn pushes us toward lower boron fraction.

Second, we see that the requirement of sustained burn naturally pushes us toward ion temperatures above 150 keV.
This helps to determine the minimal invested energy needed to ignite a pB11 fusion plasma, given by
\begin{align}
	\mathcal{E}_\text{in} = \sum_s U_s V. \label{eq:Ein}
\end{align}
Here, $U_s \equiv \frac{3}{2} n_s T_s$, $n_s$ is the number density of species $s$, and $V$ is the hotspot volume.

In order to determine the plasma performance, we must compare this invested energy to the output energy\cite{Wurzel2022ProgressFusion}
\begin{align}
	\mathcal{E}_\text{out} = \mathcal{E}_F n_{b,\text{burn}} V. \label{eq:Eout}
\end{align}
Here, $\mathcal{E}_F = 8.7$ MeV is the fusion energy released in each reaction and $n_{b,\text{burn}}$ is the density of the boron that burned.
From the input energy and output energy, we can then calculate the scientific $Q_\text{sci} \equiv  \mathcal{E}_\text{out}/ \mathcal{E}_\text{in}$.
For a reactor breakeven, we need $Q_\text{sci} \eta > 1$, where $\eta$ characterizes the efficiency of (a) turning electric energy into energy at stagnation, and (b) turning thermal energy released by the reaction into fuel. 
A 50\% efficiency in either of these would be fairly optimistic, so $\eta < 1/4$ represents a fairly optimistic assumption, and thus we need $Q_\text{sci} > 4$ to have a hope at inertial fusion energy release.

Combining equations (\ref{eq:Ein}-\ref{eq:Eout}), and invoking quasineutrality ($n_e = \sum_i Z_i n_i)$, and assuming $n_b \leq n_p$, we can derive an expression for $Q_\text{sci}$ in terms of a few parameters of the reaction:
\begin{align}
	Q_\text{sci} = \frac{2}{3} \mathcal{B} \frac{\mathcal{E}_F}{T_i} \Phi_b , \; \mathcal{B} \equiv \left(\frac{f_b}{1 + \bar{T}_e \left[1+(Z_b - 1) f_b\right]} \right). \label{eq:QsciDerived}
\end{align}
Here, $f_b = n_{b0}/(n_{p0} + n_{b0})$ is the initial boron fraction, $\Phi_b = n_{b,\text{burn}} / n_{b0}$ is the boron burnup fraction, and $\bar{T}_e = T_e/T_i$ is the initial ratio of electron temperature to ion temperature.
The last equality involved calculating the electron density using quasineutrality. 

Eq.~(\ref{eq:QsciDerived}) shows that it is advantageous to invest as little energy in the electrons as possible ($\bar{T}_e = 0$.)
It also shows that, absent burn considerations, it is advantageous to both increase boron fraction $f_b$ (which increases $\mathcal{B}$), and decrease the initial ion  temperature.
However, the requirements of sustained burn showed us that $f_b$ and $T_i$ are coupled (Fig.~\ref{fig:TePbPf_Vs_Ti}).
Thus, taking $T_i = 200$ keV, $f_b = 0.3$, we find that when the electrons and ions are the same temperature ($\bar{T_e} = 1$):
\begin{align}
	Q_\text{sci} = 2.7 \Phi_b.
\end{align}
This means that even if all the boron burned, the plasma could not achieve breakeven.
(For context, the best-performing current shots on the National Ignition Facility (NIF) have a burnup fraction of 4.4\%.\cite{Callahan2024ProspectusLaserdriven})

A somewhat more optimistic result occurs if one assumes that the electrons are kept cold during the compression ($\bar{T_e} = 0$), reducing the invested energy.
However, even then, we only have:
\begin{align}
	Q_\text{sci} = 8.7 \Phi_b.
\end{align}
Furthermore, accessing this regime assumes that we can still achieve burn at the same $T_i$ if we start with cold electrons.

Finally, it must be noted that the above estimates are optimistic, in the sense that as the burn proceeds, the fusion rate will be reduced relative to the bremsstrahlung rate by virtue of the diminishing fuel volumes, making high burnup hard to achieve.
It seems that the initial temperatures to produce a burning plasma in the presence of bremsstrahlung losses are simply too high to allow for power plant breakeven.

However, in arriving at this conclusion, we made a major assumption: that bremsstrahlung radiation is lost from the plasma.
If some fraction of the bremsstrahlung is reabsorbed through inverse bremsstrahlung, then the plasma might be able to achieve sustained burn at lower ion temperature.
Indeed, such inverse bremsstrahlung processes are thought to be an important determinant of the burn in DT plasmas.\cite{Hurricane2019ApproachingBurning}

Thus, in the next sections, we will develop the burn model without bremsstrahlung reaborption, confirming that breakeven is impossible in this case.
Then, we will develop an inverse bremsstrahlung absorption model, and incorporate it into the rate equations, showing how this bremsstrahlung absorption ultimately can allow for power plant gain.

\section{Burn model} \label{sec:BurnModel}

The pB11 fusion reaction can be simply modeled by a set of coupled rate equations describing the change in particle density $n_s$ of the protons $p$, boron $b$, and helium $\alpha$, and the change in the energy density $U_s \equiv \tfrac{3}{2} n_s T_s$ of $\alpha$'s, fuel ions $i$, and electrons $e$: 
\begin{align}
	\frac{dn_\alpha}{dt} &= 3 K_F f_\alpha \label{eq:dnadt} \\
	\frac{dn_p}{dt} &= - K_F  \label{eq:dnpdt}\\
	\frac{dn_b}{dt} &= - K_F  \label{eq:dnbdt}\\
	\frac{dU_\alpha}{dt} &= K_F \left(\mathcal{E}_F + 3 T_i \right) +   \sum_{s \neq \alpha} K_{\alpha s} (T_s - T_\alpha)  \label{eq:dUadt}\\
	\frac{dU_i}{dt} &=  - 3 K_F T_i   + \sum_{s \neq i} K_{i s}  (T_s - T_i)  \label{eq:dUidt}\\
	\frac{dU_e}{dt} &= - P_B + \sum_{s \neq e} K_{e s}  (T_s - T_e) . \label{eq:dUedt}
\end{align}
We also assume quasineutrality, i.e. $n_e = \sum_j Z_j n_{j}$, for $j \in \{p,b,\alpha\}$.
Here, $P_H$ is the external heating power, and $\mathcal{E}_F = 8.7$ MeV is the energy released in the fusion reaction.
The $K_{i j}$ are rate constants of energy transfer collisions between species $i$ and $j$, related to the thermalization collision frequencies $\nu_{ij}$ by $K_{i j} = \frac{3}{2} \nu_{ij} n_i$, which is symmetric in $i$ and $j$ as $\nu_{ij} \propto n_j$.\cite{Ochs2022ImprovingFeasibility}
Consistent with Refs.~\onlinecite{Ochs2022ImprovingFeasibility,Ochs2024ErratumImproving}, $P_F$ and $P_B$ represent the fusion and bremsstrahlung power densities respectively.
$P_F$ uses the Sikora-Weller cross sections,\cite{Sikora2016NewEvaluation} accurate to within $\lesssim 3.5\%$, multiplied by an enhancement factor\cite{Ochs2022ImprovingFeasibility,Ochs2024ErratumImproving} to effectively capture kinetic broadening of the ion tails due to collisions with hot helium.
$P_B$ uses the fit from Ref.~\onlinecite{Svensson1982ElectronPositronPair}, and has been shown to be accurate to within $\lesssim 2\%$ in the relativistic regime considered here.\cite{Xie2024BremsstrahlungRadiation}

To initialize a simulation, we set the total ion density $n_i = n_p + n_b$, the boron fraction $f_b = n_b / n_i$, the ion temperature $T_i$, and the electron-to-ion temperature ratio $\bar{T}_e = T_e / T_i$.
There is also one more variable that we set; the areal density electron density of the implosion $n_e R$, where $R$ is the plasma radius.
This parameter sets the total burn time (end point of the simulation), which is given by\cite{Rosen1999PhysicsIssues}
\begin{align}
	T_\text{burn} = \frac{R}{4 C_s},
\end{align}
where $C_s$ is the sound speed.
In practice, as we also consider implosions with low electron temperature relative to ion temperature, we instead take 
\begin{align}
	T_\text{burn} = \frac{R}{4 \times \text{max} ( C_s, v_{\text{th},p})},
\end{align}
where $v_{\text{th},p}$ is the proton thermal velocity.
For simplicity, we consider only the initial temperatures and densities when estimating the burn time.
Then, with these parameters specified, Eqs.~(\ref{eq:dnadt}-\ref{eq:dUedt}) are solved using the BDF method as implemented in \texttt{scipy.interpolate.solve\_ivp}.

Finally, the simulation proceeds until either (a) the burn time is reached, (b) the plasma becomes highly coupled (i.e. the Coulomb logarithm in the electron-ion collision frequency reaches 1), or (c) the electron temperature falls below 5 keV.
The last condition is included for consistency with later bremsstrahlung absorption simulations, and to avoid numerical issues.
For each simulation, we check whether the burn has completed, which we define as the final burn rate being less than 10\% of the maximum burn rate.

In the results that follow, for each set of $n_i $,  $T_i$, $\bar{T}_e$, and $n_e R$, we first scan over $f_b$ to find the optimum proton-boron mix.
Because the curve $Q_\text{sci}(f_b)$ is relatively smooth, we find the optimum by performing a spline fit for $Q_\text{sci}(f_b)$ and taking the maximum of the spline interpolator function to yield $f_b^*$.
Other quantities of interest, such as the burn fraction $\Phi_b$, can then be fit by evaluating their own spline interpolator functions at $f_b^*$.

Finally, at the termination of each simulation, we can keep track of whether the burn completes, which we define as the condition that the final burn rate is equal to at least 10\% of the maximum burn rate.
This metric reveals whether the plasma might have achieved better performance with a longer burn, i.e. at higher areal density $n_e R.$

\section{Maximum $Q_\text{sci}$ without bremsstrahlung reabsorption} \label{sec:QTransparent}

\begin{figure}
	\centering
	\includegraphics[width=\linewidth]{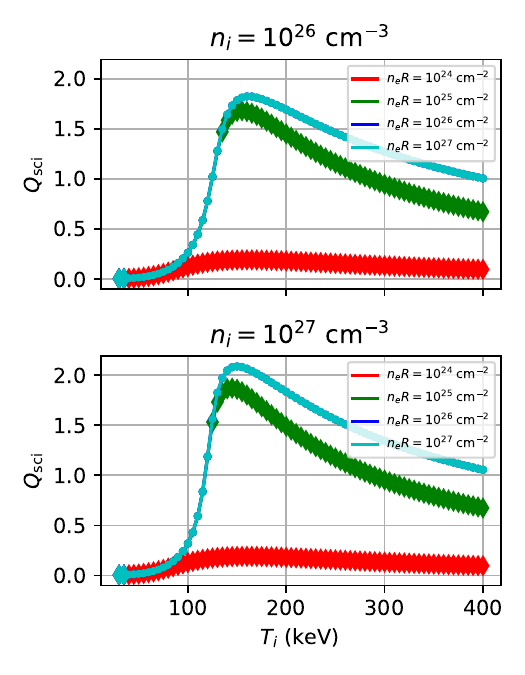}
	\caption{$Q_\text{sci}$ as a function of $T_i$ for a plasma transparent to bremsstrahlung, for equal initial electron and ion temperatures ($\bar{T}_e = 1$), and for several values of the ion density $n_i$ and electron areal density $n_e R$.
	Dots indicate simulations where the burn completed ($P_{F,\text{final} }< P_{F,\text{max}}/10$), and diamonds indicate simulations where the burn was still ongoing when the simulation ended at the disassembly time.}
	\label{fig:QSci_Transparent}
\end{figure}

\begin{figure}
	\centering
	\includegraphics[width=\linewidth]{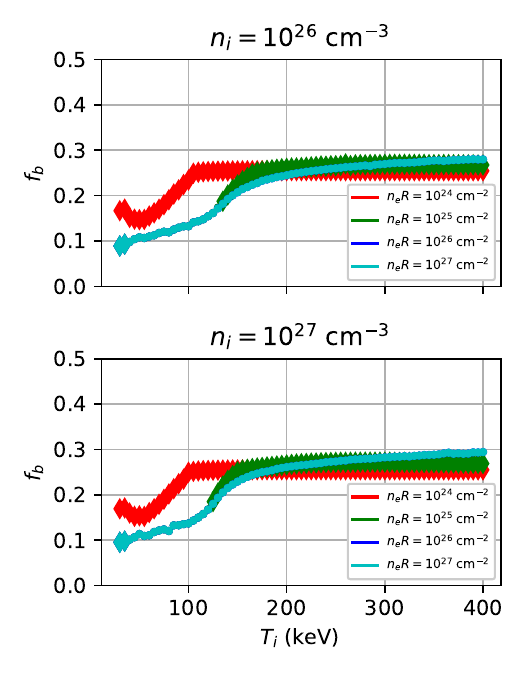}
	\caption{Optimal boron fraction $f_b \equiv n_{b0} / n_{i0}$ for the simulations without bremsstrahlung reabsorption [Figs.~\ref{fig:QSci_Transparent}-\ref{fig:QSci_Transparent_LowTe}]. In line with the discussion in Sec.~\ref{sec:heuristics}, $f_b \sim 0.25$ seems to represent the optimal tradeoff between limiting bremsstrahlung radiation and burning as much fuel as possible.
	As before, dots indicate completed burn, while diamonds indicate incomplete burn.}
	\label{fig:BoronFraction_Transparent}
\end{figure}

\begin{figure}
	\centering
	\includegraphics[width=\linewidth]{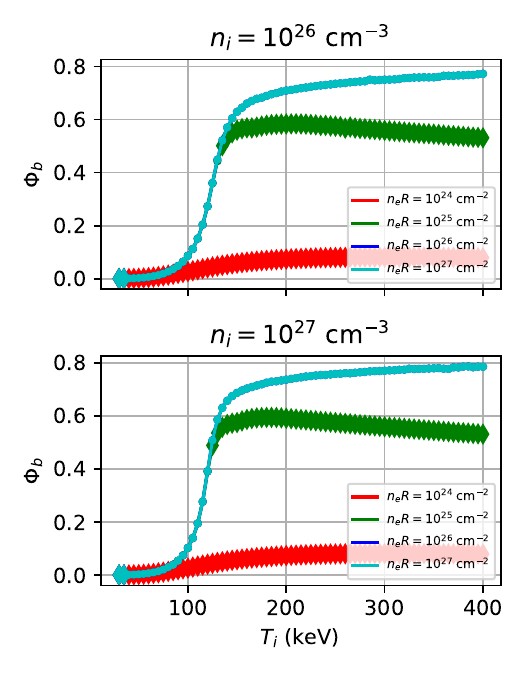}
	\caption{Boron burn fraction $\Phi_b \equiv n_{b,\text{burn}}/n_{b0}$ for the simulations without bremsstrahlung reabsorption [Figs.~\ref{fig:QSci_Transparent}-\ref{fig:QSci_Transparent_LowTe}]. 
	As before, dots indicate completed burn, while diamonds indicate incomplete burn.
	The burn fraction hovers between 60-70\% for the optimal cases.}
	\label{fig:BurnFraction_Transparent}
\end{figure}

\begin{figure}
	\centering
	\includegraphics[width=\linewidth]{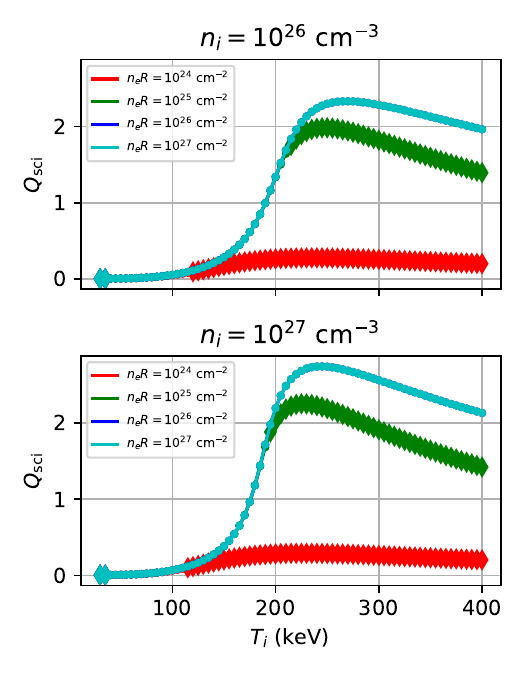}
	\caption{$Q_\text{sci}$ as a function of $T_i$ for a plasma transparent to bremsstrahlung, as in Fig.~\ref{fig:QSci_Transparent}, but now with an initial electron temperature equal to 1/5 the initial ion temperature ($\bar{T}_e = 0.2$).
	As before, dots indicate completed burn, while diamonds indicate incomplete burn.}
	\label{fig:QSci_Transparent_LowTe}
\end{figure}

The results of the burn simulations without bremsstrahlung reabsorption are shown in Figs.~\ref{fig:QSci_Transparent}-\ref{fig:QSci_Transparent_LowTe}.
To start, we assume that initially $T_e = T_i$, and we examine $Q_\text{sci}$ as a function of $T_i$ for several values of the electron density and areal density [Fig.~\ref{fig:QSci_Transparent}].
First, and most importantly, it is immediately clear from these simulations that, as predicted in Sec.~\ref{sec:heuristics}, it is hard to get above a $Q_\text{sci} \sim 2$, which occurs at an optimum initial ion temperature around 160 keV.

Second, it is clear that near-complete burn requires an areal density a bit above $n_e R = 10^{25}$ cm$^{-2}$, regardless of the value of $n_i$.
This areal density ensures that there is enough burn time to reach a relatively high burn fraction.

Third, we can observe that there is slightly better performance at higher $n_i$.
This performance increase is due to the reduction in the Coulomb logarithm at higher density, which reduces ion-electron equilibration and bremsstrahlung rates relative to fusion power. 
However, this scaling is weak compared to the effect of the areal density scaling.

The poor plasma performance is in line with our estimates from Sec.~\ref{sec:heuristics}.
The optimum is around $T_i^* = 160$ keV  and $f_b^* = 1/4$, reflecting the tradeoff in ion temperature and initial boron fraction of the fuel mix [Fig.~\ref{fig:BoronFraction_Transparent}].
Combined with the fact that the burn fraction in this regime hovers in the range of 60-70\% [Fig.~\ref{fig:BurnFraction_Transparent}], this means that $Q_\text{sci}$ struggles to reach even a modest value of 2.

Most surprisingly, perhaps, it turns out that lowering the initial electron temperature doesn't substantially help.
The lower initial electron temperature necessitates a higher initial ion temperature to sustain the burn, increasing the invested energy.
Thus, at a value of $\bar{T}_e = 0.2$, one can reach modestly higher values of $Q_\text{sci} \sim 2.2-2.8$ (Fig.~\ref{fig:QSci_Transparent_LowTe}), but not transformatively high values. 

\section{Bremsstrahlung Reabsorption Fraction} \label{sec:BremsReabsorption}

\begin{figure}
	\includegraphics[width=\linewidth]{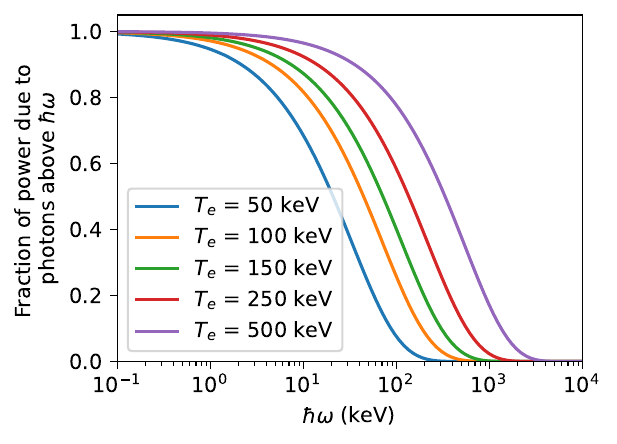}
	\caption{This figure illustrates the power dependence of the bremsstrahlung spectrum. It shows the fraction of radiated power for which the photon energy $\hbar \omega$ exceeds a given value. Equivalently, this can be understood as the fractional of radiated power that escapes the plasma if the plasma is optically thick below $\omega$ and thin above $\omega$.}\label{fig:bremsstrahlungSpectrum}
\end{figure}

At sufficiently high densities, it is natural to wonder to what extent bremsstrahlung might be reabsorbed before leaving the plasma. 
Efficient reabsorption might substantially mitigate the radiative loss channel. 
The reabsorption of radiation by plasma is strongly dependent on the frequency of the radiation. 
Figure~\ref{fig:bremsstrahlungSpectrum} shows the fraction of the electron-ion bremsstrahlung power that is carried by photons with energies above different values. 
The spectrum is calculated using the Bethe-Heitler formula with the Elwert correction factor,\cite{Elwert1939, Nozawa1998, Munirov2023} taking the electron distribution to be Maxwell-J\"uttner-distributed. 
The temperatures in the figure are defined such that the electron distribution goes like $e^{-\gamma m_e c^2 / T_e}$, where $\gamma$ is the Lorentz factor, $m_e$ is the electron mass, and $c$ is the speed of light. 
The curves in the figure take the ion charge state to be $Z = 1$. 
The upshot is that substantial power reduction requires the absorption of photons with $\hbar \omega \sim \mathcal{O}(T_e)$. 

Of course, a real plasma will smoothly transition between optically thick and thin regimes as a function of photon frequency. 
For our power balance calculations, we will use the following model for the absorption inverse scale length: 
\begin{gather}
\kappa = \nu_{ei} \frac{n_e / n_c}{c \sqrt{1 - n_e / n_c}} \, . \label{eqn:absorptionKappa}
\end{gather}
Here $\nu_{ei}$ is the electron-ion collision frequency, $n_e$ is the electron number density, and $n_c$ is the critical density, given by 
\begin{gather}
n_c \doteq \frac{m_e \omega^2}{4 \pi e^2} \, . 
\end{gather}
$\omega$ is the frequency of the radiation being absorbed. 

Eq.~(\ref{eqn:absorptionKappa}) is essentially consistent with the treatment of inverse bremsstrahlung absorption used in the ICF literature. 
Modeling this process has been a subject of active investigation in recent years.\cite{Devriendt2022, Turnbull2023, Turnbull2024}
Note, however, that there are correction factors that are used in modeling existing laser-plasma experiments that we do not include in Eq.~(\ref{eqn:absorptionKappa}). 
One of these is the Langdon effect, which accounts for non-Maxwellian features of the electron distribution in laser-heated plasmas.\cite{Langdon1980, Gonzalez2025} 
Another is the effect of collective ion screening, particularly when $n_e / n_c$ is not small.\cite{Dawson1962, Dawson1963, Rozsnyai1979, Armstrong2014, Devriendt2022} 

We take the perspective that it is better to use a relatively na\"ive absorption model than to include more sophisticated corrections that may not be appropriate for our application. 
In the case of the Langdon effect, the model was developed for the case in which the radiation comes from high-intensity laser light (acting as an external forcing term). 
Here, the radiation is instead emitted from the plasma as bremsstrahlung. 
We should expect that any radiation-driven non-Maxwellian features may look substantially different in these two cases. 
Similarly, the screening corrections used in the laser plasma literature may not be appropriate for the regime we consider here (considering the large differences in temperature and photon spectrum between what is generally seen in the laser plasma literature and what would be found in a pB11 ICF plasma). 
In any event, we will note that the characteristic sizes of these corrections is discussed in some detail in Turnbull \textit{et al.},\cite{Turnbull2023} and they are generally $\mathcal{O}(1)$ rather than, say, $\mathcal{O}(10)$. 

\begin{figure}
	\includegraphics[width=\linewidth]{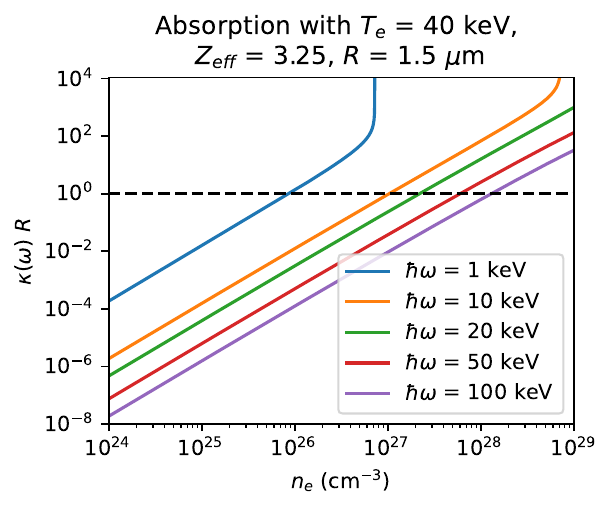}
	\caption{Variation of the radiation absorption for different frequencies at characteristic parameters and varying plasma density. Note the rapid increase in $\kappa$ as $n_e$ approaches $n_c$ (shown in the figure for the lower frequencies). The dashed line marks $\kappa R = 1$. } \label{fig:densityKappa}
\end{figure}

The absorption will be set by the dimensionless parameter $\kappa R$, where $R$ is the system size. 
Consider the behavior of $\kappa R$ as given by Eq.~(\ref{eqn:absorptionKappa}). 
$\kappa R$ scales quadratically with the ion charge $Z$. 
Neglecting logarithmic corrections, $\kappa R \propto T_e^{-3/2}$. 
Moreover, so long as $n_e \ll n_c$, $\kappa R$ scales linearly with $n_e^2 R$ (where $R$ is the system size). 
If $n_e$ more closely approaches $n_c$, $\kappa R$ rapidly increases. 
This rapid increase is shown in Figure~\ref{fig:densityKappa}. 

\begin{figure}
    \includegraphics[width=\linewidth]{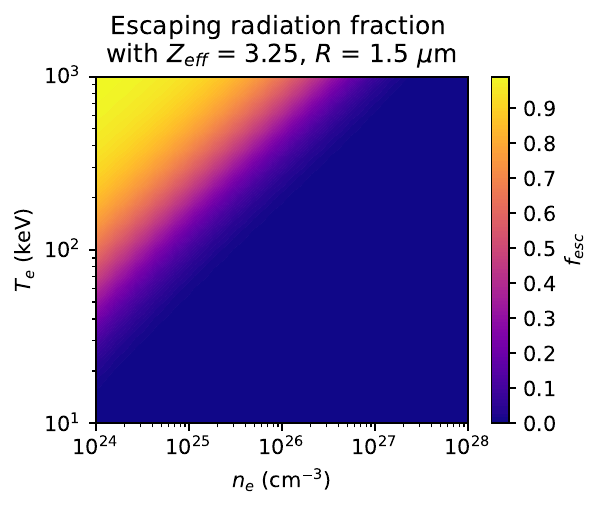}
	\caption{The fraction of bremsstrahlung power that escapes the plasma as a function of density and temperature. } \label{fig:fescPlot}
\end{figure}

\begin{figure}
	\includegraphics[width=\linewidth]{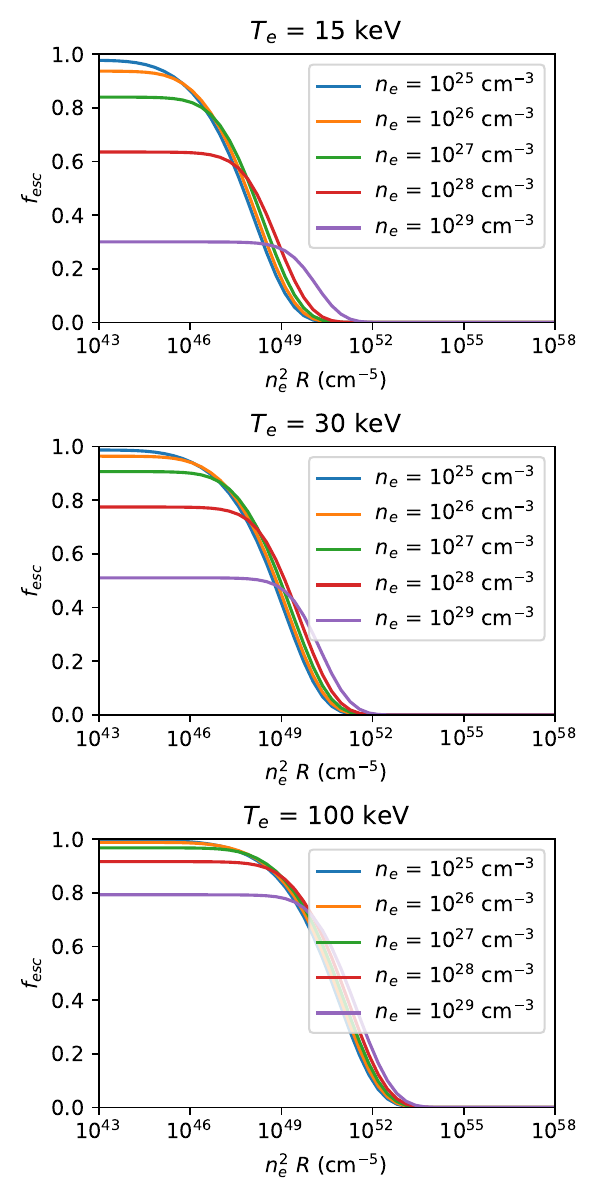}
	\caption{Dependence of the radiation escape fraction on $n_e^2 R$ with $Z_{eff} = 3.25$. For higher densities and temperatures, $f_\text{esc}$ depends on $n_e$ almost entirely through the combination $n_e^2 R$. } \label{fig:neSquaredR}
\end{figure}

The radiation loss can be modeled as follows: 
\begin{gather}
P_B = \int_0^\infty \frac{\D P_{B,\text{emitted}}}{\D \omega} \, e^{-\kappa(\omega) R} \, \D \omega. 
\end{gather}
Here $P_B$ is the Bremsstrahlung power loss that actually escapes the plasma, whereas $P_{B,\text{emitted}}$ is the power loss rate without absorption (for example, as described by Eq.~(8) in Munirov and Fisch\cite{Munirov2023}). 
The attenuation factor $e^{-\kappa R}$ for any given frequency $\omega$ very quickly approaches zero when $\kappa(\omega) R$ increases past 1, and very quickly approaches 1 when $\kappa(\omega) R$ becomes small. 
The fraction of radiated power that escapes will be denoted by $f_\text{esc}$: 
\begin{gather}
f_\text{esc} \doteq \frac{P_B}{P_{B,\text{emitted}}} \, . \label{eq:fEsc}
\end{gather}
Figure~\ref{fig:fescPlot} shows the variation of $f_\text{esc}$ with density and temperature. 
Note that increasing density tends to decrease $f_\text{esc}$, whereas increasing temperature has the opposite effect. 
Figure~\ref{fig:neSquaredR} shows the dependence on $n_e^2 R$. 
The regimes in which $f_\text{esc}$ depends on $n_e$ independently of $n_e^2 R$ illustrate the importance of the radiation cutoff when $n_e$ approaches or exceeds $n_c$ for a given range of frequencies. 

\subsection{Incorporating Reabsorption in the Burn Model}  \label{sec:BurnModelBremsReabsorption}

To incorporate the bremsstrahlung radiation reabsorption robustly into the burn model, we simply multiply the bremsstrahlung power $P_B$ in Eq.~(\ref{eq:dUedt}) by the factor $f_\text{esc}$ from Eq.~(\ref{eq:fEsc}).
Because the calculation of $f_\text{esc}$ is fairly expensive, we evaluated the function at equally-spaced points in the space of $[\log_{10} n_e, \log_{10} (n_e^2 R), T_e, Z_\text{eff}]$.
We then evaluated the points using the fast linear interpolator function \texttt{map\_coordinates()}, implemented in the \texttt{scipy.ndimage} library.
The points we took to build this 4D interpolator were: (a) $\log_{10} n_e$ spaced from 25 to 29 $\log_{10}$ (cm$^{-3})$ in increments of 1; (b) $\log_{10} n_e^2 R$ spaced from 43 to 59 $\log_{10}$ (cm$^{-5})$ in increments of 1; (c) $T_e$ spaced from 5 keV to 420 keV in increments of 5 keV; and (d) $Z_\text{eff}$ spaced from 1 to 5 in increments of 1.

Unfortunately, depending on the plasma density and electron temperature, the formula for $f_\text{esc}$ can exhibit a few divergences that must be handled.
The simulation stop conditions generally prevent incursion into this problematic regime, but to prevent numerical issues with the spline due to divergences, we made a couple corrections to the formula for $f_\text{esc}$.
First, we took the Coulomb logarithm to have a minimum value of 1, preventing incursion into the strong-coupling regime.
Second, we assumed zero emission above the critical density (i.e. frequencies below the plasma frequency).
It should be emphasized that these changes should not affect the numerical results, which are terminated when plasma becomes strongly coupled or degenerate.

\section{Maximum $Q_\text{sci}$ with bremsstrahlung reabsorption} \label{sec:QWithReabsorption}

\begin{figure}
	\centering
	\includegraphics[width=\linewidth]{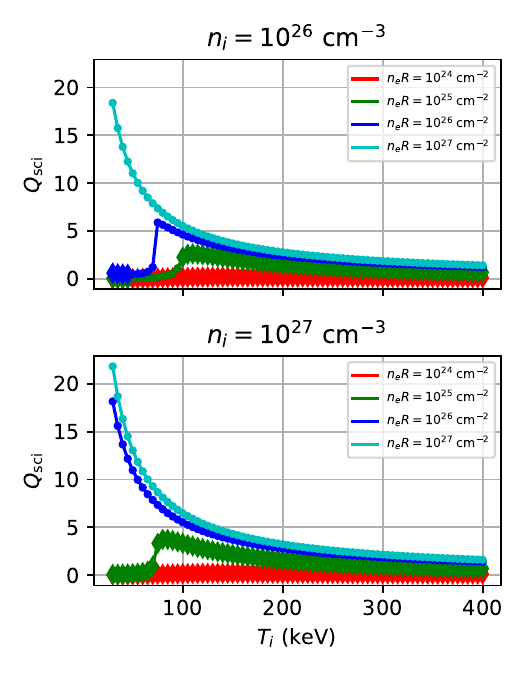}
	\caption{$Q_\text{sci}$ as a function of $T_i$ for a plasma, for the same parameters as in Fig.~\ref{fig:QSci_Transparent}), but now accounting for bremsstrahlung reabsorption. Here, we take equal initial electron and ion temperatures ($\bar{T}_e = 1$).
	As before, dots indicate completed burn, while diamonds indicate incomplete burn.
	In contrast to the simulations without bremsstrahlung reabsorption, the achievable $Q_\text{sci}$ is much higher, especially at low ion temperature.}
	\label{fig:QSci_Abs}
\end{figure}

\begin{figure}
	\centering
	\includegraphics[width=\linewidth]{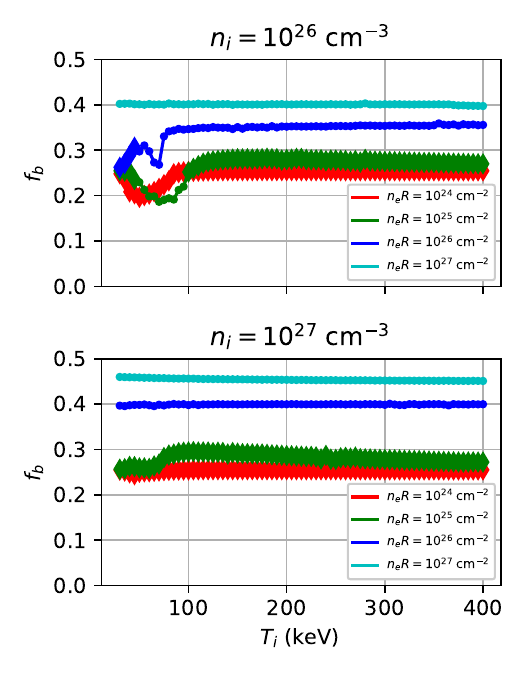}
	\caption{Optimal boron fraction $f_b \equiv n_{b0} / n_{i0}$ for the simulations with bremsstrahlung reabsorption [Figs.~\ref{fig:QSci_Abs}-\ref{fig:BurnFraction_Abs}]. 
	As before, dots indicate completed burn, while diamonds indicate incomplete burn.
	In contrast to the case without bremsstrahlung reabsorption [Fig.~\ref{fig:BoronFraction_Transparent}], the boron fraction for high-performing plasma is higher in plasmas with high degrees of bremsstrahlung reabsorption.}
	\label{fig:BoronFraction_Abs}
\end{figure}

\begin{figure}
	\centering
	\includegraphics[width=\linewidth]{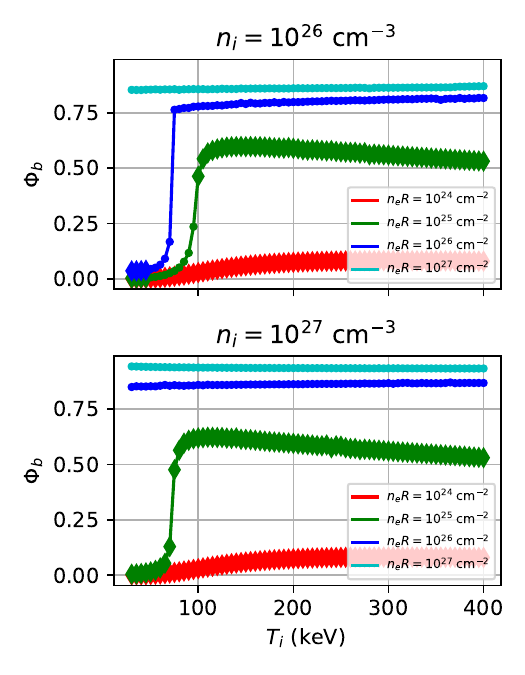}
	\caption{Boron burn fraction $\Phi_b \equiv n_{b,\text{burn}}/n_{b0}$ for the simulations without bremsstrahlung reabsorption [Figs.~\ref{fig:QSci_Abs}-\ref{fig:BurnFraction_Abs}].
	As before, dots indicate completed burn, while diamonds indicate incomplete burn. 
	The burn fraction is above 80\% for the optimal cases.}
	\label{fig:BurnFraction_Abs}
\end{figure}

The results of the burn simulations are shown in Figs.~\ref{fig:QSci_Abs}-\ref{fig:BurnFraction_Abs}.
The case for $T_e = T_i$ is shown in Fig.~\ref{fig:QSci_Abs}.
The main conclusion, immediately clear from the figure, is that bremsstrahlung absorption allows the plasma to reach much higher values of $Q_\text{sci} \sim 15-20$. 
Furthermore, these high values of $Q_\text{sci}$ now occur at much lower initial ion temperatures than before, suggesting that an ICF plasma with bremsstrahlung absorption favors a slower, lower-temperature burn.
Such an optimum makes sense, since both the initial invested energy and the bremsstrahlung losses scale strongly with the electron temperature.

Second, we can observe a change in the scaling of $Q_\text{sci}$ with $n_e R$.
In the case without bremsstrahlung absorption, $Q_\text{sci}$ basically depended on the density only through the areal density $n_e R$, with only a mild dependence on the density itself.
However, because the bremsstrahlung reabsorption depends on the density primarily through $n_e^2 R$, the parameter $n_e R$ alone is no longer sufficient to determine the plasma performance. 
This new scaling is most apparent comparing the blue lines (electron density $n_e R = 10^{26}$ cm$^{-2}$ in Fig.~\ref{fig:QSci_Abs}): it is clear that at the higher density (and thus higher $n_e^2 R$) the plasma is able to achieve higher values of $Q_\text{sci}$ due to the greater bremsstrahlung reabsorption.

Third, we can observe that in plasmas with a high degree of bremsstrahlung reabsorption, the optimal boron fraction $f_b$ is substantially higher than in the simulations without bremsstrahlung reabsorption [Fig.~\ref{fig:BoronFraction_Abs}].
This new optimum reflects the fact that boron is not as costly when bremsstrahlung losses are limited by reabsorption, and thus a more equal fuel mix is desirable.

Finally, we see that the burn fraction $\Phi_b$ is higher in plasma with substantial bremsstrahlung reaborption, since the reaction does not quench as quickly [Fig.~\ref{fig:BurnFraction_Abs}].

Taken together, these results show that bremsstrahlung reabsorption is a necessary condition to achieve high plasma performance in an ICF pB11 fusion plasma.

\section{Implications for Yield} \label{sec:Yield}

We have identified two rough constraints on the plasma density and radius in order to achieve high performance.
First, we need a sufficiently long burn, which puts a constraint on the areal density $n_e R \gtrsim 3\times 10^{25}$ cm$^{-2}$.
For a 60-40 proton-boron mix, this corresponds to an areal mass density of 100 g/cm$^{2}$.
Second, we need the plasma to be sufficiently optically thick to bremsstrahlung, which requires $n_e^2 R \gtrsim 10^{53}$ cm$^{-5}$.
The latter condition will be more restrictive at low densities $n_e \lesssim 3 \times 10^{27}$ cm$^{-3}$ (i.e. $n_i \gtrsim 10^{27}$ for a 60-40 proton-boron mix), and the former at higher densities $n_e \gtrsim 3 \times 10^{27}$ cm$^{-3}$.

However, there is an additional criterion we need to meet: reasonable yields given laser compression technology.
Currently planned lasers go up to 10 MJ, which for a $Q_\text{sci} \sim 10$ plasma would imply a 100 MJ yield.
Assuming complete burn, the yield scales as $Y = 4\pi  n_b \mathcal{E}_F R^3 / 3$.
This can be expressed in terms of $n_e^2 R$ and $n_e R$.
\begin{align}
	Y = \frac{4\pi}{3} \frac{n_b}{n_e^3} (n_e R)^3 \mathcal{E}_F = \frac{4\pi}{3}  \frac{n_b}{n_e^6} (n_e^2 R)^3 \mathcal{E}_F.
\end{align}

Combining the constraints of $Y \lesssim 10^8$ J, $n_e R \gtrsim 3 \times 10^{25}$ cm$^{-2}$, and $n_e^2 R \gtrsim 10^{53}$ cm$^{-5}$, we arrive at:
\begin{align}
	n_e &\gtrsim \max \biggl[ \left(\frac{4\pi}{3} \frac{n_b}{n_e} \frac{(8.7 \text{ MeV})(3 \times 10^{25} \text{ cm}^{-2})^3 }{100 \text{ MJ}}  \right)^{1/2}, \notag\\
	& \hspace{0.5in} \left(\frac{4\pi}{3} \frac{n_b}{n_e} \frac{(8.7 \text{ MeV})(1 \times 10^{53} \text{ cm}^{-5})^3 }{100 \text{ MJ}}  \right)^{1/5} \biggr],
\end{align}
where the first line comes from the areal density constraint, and the second line from the bremsstrahlung reabsorption constraint.
For a 60-40 proton-boron mix, $n_b/n_e = 0.15$. 
The areal density constraint dominates, and we find $n_e \gtrsim 1.5 \times 10^{28}$ cm$^{-3}$, i.e. $n_i \gtrsim 6 \times 10^{27}$ cm$^{-3}$.

\begin{figure}
	\centering
	\includegraphics[width=\linewidth]{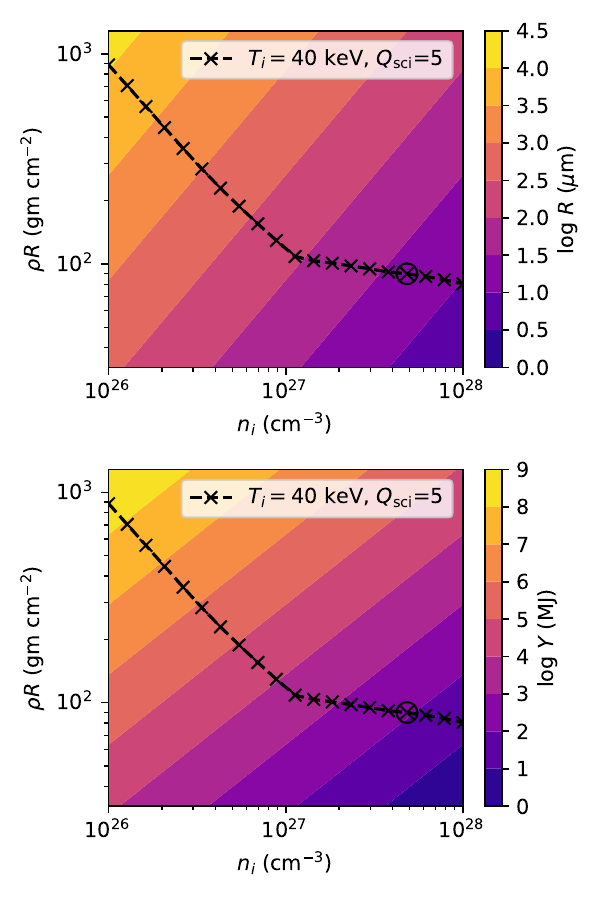}
	\caption{Plasma radius $R$ (top) and total yield at 100\% burn (bottom) as a function of initial fuel ion density $n_i$ and areal density $\rho R$ for a proton-boron ICF fusion plasma with a 60-40\% proton-boron mix.
	Superimposed is a set of simulations (marked $\times$) showing the necessary conditions to achieve $Q_\text{sci}=5$ for a plasma with an initial ion temperature of 40 keV.
	The kink in the curve represents the transition between the areal density being determined by the bremsstrahlung capture constraint ($n_e^2 R \lesssim 10^{53}$ cm$^{-5}$), vs. the burn time constraint  ($n_e R \lesssim 3 \times 10^{25}$ cm$^{-2}$); as predicted, this kink occurs at an ion density of $n_i \sim 10^{27}$ cm$^{-3}$, resulting in a minimal areal density of 100 gm/cm$^{2}$.
	Although this is not a complete optimization, it can be seen that in order to achieve $Q>5$ at a tolerable yield ($\sim 100$ MJ), it is necessary to achieve ion densities above $6 \times 10^{27}$ cm$^{-3}$, and areal densities around 100 gm/cm$^{2}$.
	These numbers are roughly two orders of magnitude above current NIF parameters, and thus not realistic in the near future.}
	\label{fig:Yield.pdf}
\end{figure}

To check these estimates more systematically, we can explicitly calculate both the plasma radius and the yield (assuming 100\% burnup and a 60-40 proton-boron mix) as a function of the ion density $n_i$ and areal density $\rho R$.
Then, we can run a series of simulations, assuming initial 40 keV ions and electrons, for several values of $n_i$.
Using a root finder(\texttt{scipy.optimize.root}), we can then find the minimal areal density $\rho R = \rho R^*$ required to achieve $Q_\text{sci} = 5$.
This results in a curve $\rho R^*(n_i)$, which can be overlaid on the contour plots of the plasma radius and yield.
The point where this curve intersects the 100 MJ yield contour represents the minimum target density for such a high-performance plasma.

The result of these simulations is shown in Fig.~\ref{fig:Yield.pdf}.
It can be seen that to get to power plant values, one needs $n_i \gtrsim 6 \times 10^{27}$ cm$^{-3}$.
Both this ion density and the required areal densities are approximately two orders of magnitude greater than current NIF hotspot parameters.\cite{Zylstra2022BurningPlasma}

\section{Discussion and Conclusion} \label{sec:Discussion}

The analysis presented here was extremely simplified.
Many of the assumptions were actually optimistic: we neglected thermal conduction losses and expansion cooling losses.
However, the assumption of uniform density does preclude certain concepts, such as fast ignition, which rely on propagating a burn wave through a very low-temperature, extremely dense plasma.
Nonuniform density arrangements, such as the typical cold shell surrounding a high-temperature hotspot, can also be helpful in increasing the areal density while keeping a lower hotspot density.
Thus, nonuniform or fast ignition plasmas might achieve dense initial conditions with lower invested energy, resulting in higher $Q_\text{sci}$.
However, it is important to note that the degeneracy pressure of the plasma makes it hard to make the density much higher at lower temperatures; at $n_i = 10^{28}$ cm$^{-3}$, the Fermi temperature is already over 30 keV.
All together, it is hard to escape the main conclusion of the paper: that 
the pB11 reaction must occur in unprecedentedly dense ICF regimes, likely requiring at least next-generation lasers.\cite{Thomas2024HybridDirect}

\begin{figure}
	\centering
	\includegraphics[width=\linewidth]{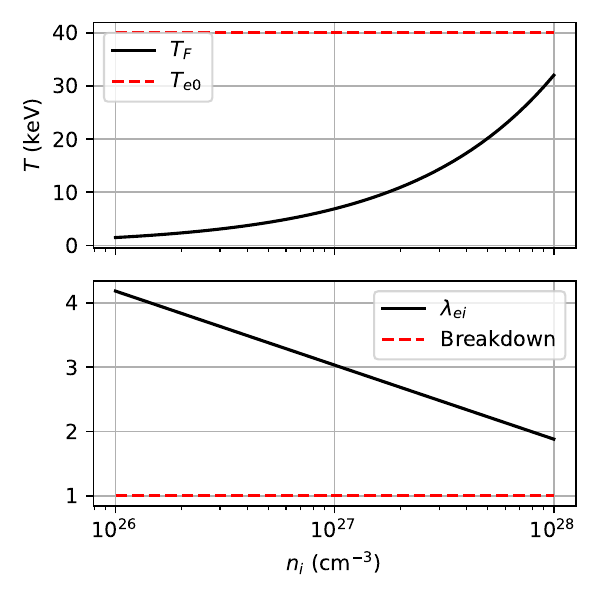}
	\caption{Fermi temperature (top) and Coulomb logarithm for a 40 keV plasma (bottom) as a function of ion density for a 60-40 proton-boron mix plasma.
	As the plasma gets more dense, the increase in the Fermi temperature and decrease in the Coulomb logarithm lead to breakdowns in the ideal plasma and collision theory, which assume nondegenerate electrons and weak coupling.}
	\label{fig:DegeneracyAndCoupling}
\end{figure}

The most major uncertainty of the current work relates to the breakdown of the model in the most optimistic regime.
Figs.~\ref{fig:QSci_Abs} and \ref{fig:Yield.pdf} show that the combined demands of high $Q_\text{sci}$ and low yield push ICF pB11 naturally toward a low-temperature, high-density regime.
Indeed, it appears from Fig.~\ref{fig:QSci_Abs} that the lower the temperature the better the performance (due to the lower invested energy).
However, in Fig.~\ref{fig:Yield.pdf}, we chose 40 keV, rather than going to aggressively lower temperatures.
The reason is that as we get to these high densities and low temperatures, the electron degeneracy physics becomes important, and multiple aspects of the rate equation model in Sec.~\ref{sec:BurnModel} break down; namely, the ideal gas law for the electrons, the electron-ion collision operator, and the bremsstrahlung operator.
For 40 keV electrons, we can see in Fig.~\ref{fig:DegeneracyAndCoupling} that both the degeneracy parameter (the ratio of electron temperature to the Fermi temperature) and the Coulomb logarithm approach one as the density approaches $n_i \sim 10^{28}$ cm$^{-3}$, both signs that the electron models underlying the burn model are beginning to break down.

In summary, then, the requirements of high $Q_\text{sci}$---i.e. bremsstrahlung reabsorption and long burn time---combined with the requirement of low yield naturally pushes an ICF proton-boron plasma toward a regime where electron degeneracy physics becomes important.
This is not necessarily a bad outcome, since both the electron-ion collisions and bremsstrahlung emission tend to be reduced in this regime.\cite{Son2006IgnitionRegime,Son2006ControlledFusion,Nazirzadeh2017InvestigationInertial}
Thus, further work into pB11 ICF should start by incorporating equation of state and collision models appropriate to this regime, as well as effects on the fusion reactivity, such as Salpeter screening,\cite{Salpeter1954ElectronScreening} which might occur in this regime.
However, it should be noted that just because we cannot rule out the degenerate regime on the basis of the calculations put forth here does not mean that the regime is not still extremely challenging to access for an extended stagnation period.

Finally, an issue that any pB11 reactor concept has to address is the issue of the neutronic He-B side-chain reaction.\cite{Walker1949CrossSection,Bonner1956NeutronsGamma,Liu2020MeasurementAlpha,Hora2021EliminationSecondary}
In magnetic confinement fusion, a prompt helium removal scheme has the potential to substantially limit the side chain reactions;\cite{Ochs2025PreventingAsh} however, such schemes are not available to the hot, dense plasmas characteristic of ICF. 
Thus, an ICF pB11 plasma will always produce some irreducible level of neutrons, albeit orders of magnitude less than for deuterium-tritium plasmas.
This neutron production should be incorporated into future pB11 ICF modeling, alongside the degeneracy physics previously discussed.

\section*{Acknowledgements}



This work was supported by the Center for Magnetic Acceleration, Compression, and Heating (MACH), part of the DOE-NNSA Stewardship Science Academic Alliances Program under Cooperative Agreement DE-NA0004148, and by the National Science Foundation under Grant PHY-2308829.

\section*{Data Availability}

Data sharing is not applicable to this article as no new data were created or analyzed in this study.


\bibliography{pB11_ICF.bib, absorptionBib.bib}


\end{document}